\documentclass[useAMS, usenatbib, usegraphicx,referee,a4paper]{mn2e}
\usepackage{deluxetable}

\renewcommand{\sun}{\odot}
\newcommand{\integral}{INTEGRAL~}
\newcommand{\xmm}{\emph{XMM-Newton} }
\newcommand{\suzaku}{\emph{Suzaku} }
\newcommand{\xspec}{\emph{XSPEC} }
\newcommand{\apec}{\emph{apec} }
\newcommand{\igr}{IGR1719~}

\date{\today}

\title[X-ray Observations of IGR~J17195-4100]{X-ray Observations of INTEGRAL Discovered Cataclysmic Variable IGR~J17195-4100}

\author[V. Girish \& K.P. Singh]{V. Girish$^1$\thanks{Email:
giri@isac.gov.in},
K.P.  Singh$^2$\thanks{Email: singh@tifr.res.in}\\
$^1$Space Astronomy Group, ISRO Satellite Centre, Airport Road,
Bangalore 560017, INDIA \\
$^2$Department of Astronomy \& Astrophysics, Tata Institute of Fundamental Research, Homi
Bhabha Road, Mumbai 400 005, INDIA
}

\graphicspath{{Figures/}}

\date{Received; in original form}

\begin{document}
\maketitle

\begin{abstract}

We present analysis of archival X-ray data obtained with the \xmm and
\suzaku for a new Intermediate Polar identified as a counterpart of an
INTEGRAL discovered $\gamma$-ray source, \emph{IGR J17195-4100}.  We report
a new period of 1053.7$\pm$12.2~s in X-rays. A new binary orbital period
of 3.52$^{+1.43}_{-0.80}$~h is strongly indicated in the power spectrum of
the time series. An ephemeris of the new period proposed as the spin period
of the system has also been obtained. The
various peaks detected in the power spectrum suggest a probable disc-less
accretion system. The soft X-rays ($<$3~keV) dominate the variability seen
in the X-ray light curves.  The spin modulation shows energy dependence
suggesting the possibility of a variable partial covering accretion column.
The averaged spectral data obtained with  \xmm EPIC cameras show a multi
temperature spectra with a soft excess. The latter can be attributed to the
varying coverage of accretion curtains.

\end{abstract}


\section{Introduction}

Cataclysmic variables (CVs) are interacting binaries with a white dwarf
accreting matter from a main sequence companion star through a Roche lobe,
with a typical orbital period of few hours (see \Citealp{warner95} for a
review).  Many of the CVs have high magnetic fields and are called as
magnetic CVs (mCVs). The mCVs are further classified as polars
\Citep[see][]{cropper90}, and Intermediate polars (IPs; see
\Citealt{patterson94},\Citealt{hellier96}) based on the strength of the
observed magnetic fields in them. Polars have very high magnetic fields
(10-80~MG) which do not allow the formation of accretion discs, but help to
synchronise the spin and orbital periods to a very high degree.
Intermediate polars have somewhat  less magnetic fields compared to polars
($\sim$ 10 MG) but sufficient to disrupt the accretion disc and  truncate
it at a certain distance from the white dwarf. The IPs typically have a
spin period of few hundreds of seconds and orbital periods of few hours and
many IPs cluster around a P$_{spin}$/P$_{orb}$ $\approx$ 0.05 - 0.15
\Citep{norton04}. The accreting matter in both polars and IPs falls on to
the white dwarf through poles along the magnetic field lines. A few IPs
show disc-less accretion, and in some cases accretion stream is seen to
flip between the two poles as was first reported in V2400 Oph by
\Citet{buckley97}. Generally the IPs are hard X-ray emitters.  However,
many IPs also show a soft X-ray excess typically of a blackbody component
in soft X-rays (\Citealt{evans07}, \Citealt{martino08},
\Citealt{anzolin09}). The soft blackbody component in IPs is usually
attributed to the presence of accretion curtains \Citep{evans07}.

The INTEGRAL/IBIS soft Gamma-ray  survey \Citep{barlow06,bird07} has
detected a total of 32 CVs till now which is close to 10 per cent of INTEGRAL
detections. These detections are identified  mainly with mCVs with 22 confirmed or
probable IPs and three polars.
IGR J17195-4100 (IGR1719 hereafter) was detected  as a source in  \integral
observations by \Citet{bird04} and was later identified with an optical
object of R$_{mag}$ $\approx$ 14.3~mag and classified as an IP  by
\Citet{masetti06} based on its optical spectrum.  \Citet{butters08}
reported candidate periods of 1842.4$\pm$1.5~s and 2645.0$\pm$4.0~s in \igr
based on RXTE observations, whereas a spin period of 18.9925$\pm$0.0006~min
(1139.55$\pm$0.04~s) and an orbital period of 4.005$\pm$0.006~h were
reported by \Citet{pretorius09} from high speed optical photometry.
\citet{masetti06} determined the
distance of \igr close to 110~pc assuming an absolute magnitude M$_V$ $\sim$
9.
The X-ray luminosity of the IP as given by
\citet{masetti06} is 3.6$\times$10$^{31}$~erg~s$^{-1}$ in the
0.5-10.0~keV energy range and 5.5$\times$10$^{31}$ erg~s$^{-1}$ in the 20-100~keV
energy range.
\Citet{yuasa10} analyzed wide-band \emph{Suzaku} spectra of seventeen IPs
with \igr being one of them.  The mass, shock temperature and spectral
parameters were derived by fitting the X-ray spectra with numerically
calculated emission models and the mass of the white dwarf in \igr was
estimated as 1.03$^{+0.24}_{-0.22}$~M$_\odot$ \citep{yuasa10}.

\section{Observations and Data Analysis}

\xmm observatory \Citep{jansen01} observed \igr on 2009 September 3 for a total
of 33.6~ks (Obs ID 0601270201). The observations were continuous and taken
with the three instruments EPIC-MOS1, MOS2 and pn (\Citealt{turner01} and \Citealt{struder01}) which provide medium resolution spectroscopy in the energy
range of 0.2-12.0~keV.  We used the latest Science Analysis Software (SAS)
version 11.0 software package for data reduction and extraction of
lightcurves and generation of spectral products. The temporal and spectral
analyses were carried out using XRONOS and \xspec version 12.7
(\Citealp{arnaud96}; \Citealp{dorman01}) packages provided in
\emph{heasoft} (Version 6.11).

\suzaku \citep{mitsuda07},  the fifth Japanese X- ray astrophysical
observatory, observed \igr on 2009 February 18-19 for a total of
$\sim$73~ks.  \suzaku observes in the energy range of 0.2-600~keV with the
help of two instruments, X-ray Imaging Spectrometer (XIS;
\citealp{koyama07}) and the Hard X-ray Detector (HXD;
\citealp{takahashi07}).  The XIS consists of four X-ray imaging charge
coupled device (CCD) cameras at the focal plane of each of the four X-ray
Telescopes (XRT; \Citealp{serlemitsos07}). The energy coverage of XIS is
similar to the \xmm energy range of 0.2-12.0~keV. The XIS2 camera stopped
working near the end of 2006, and therefore we use the data only from the
remaining three XIS cameras (XIS0, XIS1 and XIS3).  The \suzaku
observations of \igr were analysed with the latest \emph{heasoft}  tools
and CALDB version 3.1.
A log of the observations of  \igr is given in Table~\ref{tab-obs}.

\begin{table*}
\caption{Log of \xmm and \suzaku observations of \igr.
		The mean background subtracted count rate in the energy range
		of 0.3-10~keV is given.}
		\label{tab-obs}

\begin{tabular}{clcccc}
	\hline
	Observatory 	&
	Instrument	 &
	Start Time	&
	End Time &
	Exposure$^a$	 &
	Count rate \\

	 	&
		 &
	(UT)	&
	(UT) &
	(s)	 &
	(count s$^{-1}$) \\\hline

	\suzaku		& XIS	 & 2009-02-18 11:03:26 & 2009-02-19 07:27:20 & 73.4 (31.3) & 0.25 	 \\
	\xmm		& MOS1	 & 2009-09-03 06:35:37& 2009-09-03 15:55:59  & 33.6 (33.6) & 1.62 	 \\
			    & MOS2	 & 2009-09-03 06:35:37 & 2009-09-03 15:56:05 & 33.6 (33.6)& 1.59 	 \\
			    & pn	 & 2009-09-03 06:57:59 & 2009-09-03 15:51:53 & 32.0
				(32.0) & 2.55 	 \\\hline

\end{tabular}

	\tablecomments{$^a$ Useful exposures are given in brackets}
\end{table*}

The \xmm data were filtered using the standard pipeline filter criteria and
canned screening criteria using the SAS tool \emph{xmmselect} to obtain
cleaned events.  A strict criteria of PATTERN=0 and FLAG=0  were used in
addition to the standard screening criteria for further extraction and
analyses of both the timing and spectral products.  A light curve from the
source free regions was extracted in the energy range of 8-12keV from the
filtered data to check for flaring and no flaring activity was observed
during the exposure time.  Subsequently the filtered data-set were used in
all the further analysis without any temporal filter. No pile-up was
observed in the CCD during the observations. A circular region of 25$''$ was
used to extract the source light curve.
The energy spectra of the source and the background region were extracted
separately from all the three EPIC cameras using the SAS task
\emph{xmmselect}. The source spectra were extracted from a circular region
of radius 25$''$, whereas the background spectra are extracted from a
nearby source-free region from the same chip.  Spectral energy responses
were generated with \emph {rmfgen} and \emph {arfgen} tasks in SAS.  Data
from  MOS1 and MOS2 were combined together and an effective spectral
response was generated by combining the responses and effective areas of
MOS1 and MOS2.  The spectra were binned for a minimum of 50 counts per
bin.

The \suzaku data were filtered using the \emph{xselect} tool available in
\emph{heasoft} by applying the standard filtering criteria typically used
for creating the cleaned event files. The light curves were extracted from
a circular region with a radius 30$''$ with a bin time of 1~s. The
background was extracted from a nearby source free regions. The background
subtracted light curves of \igr taken from the three XIS cameras were
summed together and subjected to timing analysis.

\begin{figure}
	\includegraphics[width=1.9in, angle=270]{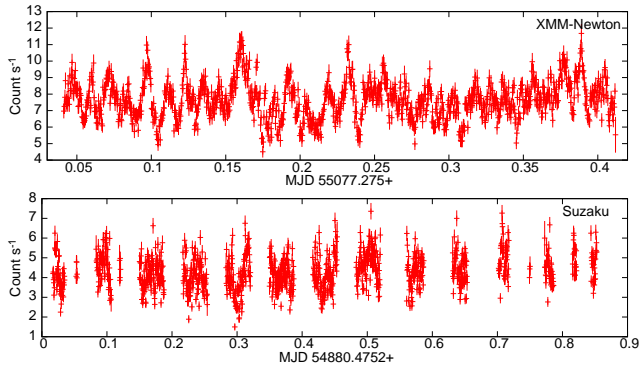}
	\caption{Broadband X-ray light curves of \igr in the energy range 0.3
	to 10.0~keV obtained from \xmm EPIC (top panel) and \suzaku XIS cameras
	(bottom panel). The light curves of different XIS and three EPIC cameras
	are summed separately. Bin time is 32~s for both the curves.}
	\label{fig-lc0}
\end{figure}

\begin{figure}
	\includegraphics[width=4.25in,angle=-90]{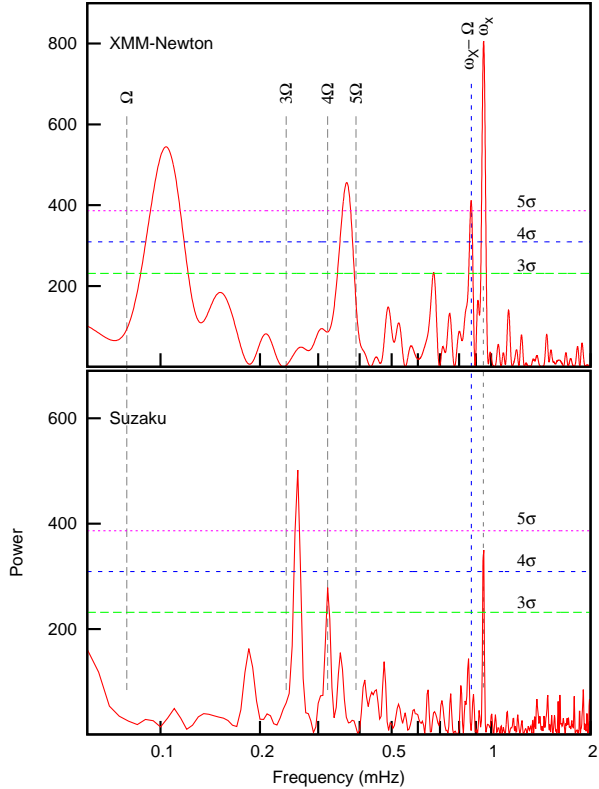}

	\caption{Expanded periodogram of \xmm (top panel) and \suzaku (bottom
	panel) data of \igr in the energy range 0.3-10.0 keV are shown along
	with the significance (3$\sigma$, 4$\sigma$, \dots) of the power peaks
	based on the standard deviation ($\sigma$) calculated from the averaged
	power spectrum in the entire range of 0-4 mHz. The strongest signal
	with the maximum power is seen at a frequency of 0.949 $\pm$ 0.011 mHz
	(1053.7 $\pm$ 12.12 s) in the \xmm data and is coincident with
	the second most dominant peak in the \suzaku data. This frequency is
	later proposed as the spin frequency of the system in the Discussion.
	The other markers correspond to the harmonics of the 
	proposed orbital period ($\Omega$=3.52~h)  of the system.
	}
	\label{fig-ft0}
\end{figure}

\subsection{ Timing Analysis }
\label{sec-timing}

The background subtracted light curves obtained from the three EPIC cameras:
MOS1, MOS2 and pn cameras in the energy range of 0.3-10.0~keV are summed
and plotted in the top panel of Figure~\ref{fig-lc0}. Similarly, the X-ray
light curve obtained from the \suzaku is shown in the bottom panel of
Fig.~\ref{fig-lc0}. The  light curves clearly show periodic intensity
variations on a time-scale of  $\sim$1000~s. The \xmm light curves are continuous
while those of \suzaku have gaps due to shadowing of the source by the
Earth. We performed a Fourier Transform of the summed \xmm light curve
using \emph{powspec} tool of FTOOLS package.  The resulting FFT power
spectrum is plotted in Figure~\ref{fig-ft0} (top panel).  The highest power
in the \xmm data corresponds to a frequency of $\omega_x$ = 0.949$\pm$0.011~mHz
(1053.7$\pm$12.2~s, Sinusoid amplitude, A$_s$ = 28.4 count s$^{-1}$).

The error quoted here corresponds $1\sigma$ errors derived from full width
at half the maximum (FWHM) of the peak. (FWHM $\sim$2.35$\sigma$ for normal
distribution).  To identify the significant peaks, we calculated the
standard deviation of the power spectrum from 0.0-4~mHz. As a conservative
approach, we included the power in all the peaks for calculating standard
deviation ($\sigma$) after removing the contribution of the highest peak in
the spectrum. The different significance levels are shown as horizontal
lines in Fig.~\ref{fig-ft0}.  There are three more peaks above 5$\sigma$
level at frequencies 0.104$\pm$0.027~mHz (9624.3$\pm$2498.6~s, A$_s$=23.7
count s$^{-1}$), 0.366$\pm$0.018~mHz (2732.2$\pm$134.4~s, A$_s$=21.3 count
s$^{-1}$) and 0.870$\pm$0.012~mHz (1149.4$\pm$15.8~s, A$_s$ = 20.3 count
s$^{-1}$) respectively in the \xmm data.  A peak very close to 3$\sigma$
level at 0.6703$\pm$0.032~mHz (1492$\pm$29~s, A$_s$ = 15.3 count s$^{-1}$)
is also seen.  A similar periodogram  of \suzaku XIS lightcurve was
obtained and it also showed several peaks.  However, that periodogram is
strongly affected by aliasing due to gaps in the \suzaku data and is
therefore not shown here.  To remove the effects of the gaps in the light
curve, we obtained a CLEAN periodogram \citep{roberts87} of XIS data, and
this is shown in Fig.~\ref{fig-ft0} (bottom panel). Since the effects of
the windowing of the data are removed, we assume the peaks seen in
Fig.~\ref{fig-ft0} as real.  The amplitudes and the probable identification
of all the significant peaks above 3$\sigma$ seen in Fig.~\ref{fig-ft0} are
listed in Table~\ref{tab-power}.

\begin{table*}
\caption{Power of various frequencies  in \igr  \label{tab-power}}
\begin{tabular}{cccrrrrrrrrc}
	\hline
	 &
	 &
	 &
	\multicolumn{9}{c}{Absolute power (count s$^{-1}$)$^2$} \\
	\cline{6-10}
		&
		 & &
	\multicolumn{4}{c}{\xmm (MOS+pn)}	 &
	 &
	\multicolumn{4}{c}{\suzaku} \\
	\cline{4-7}   \cline{9-12}
	Frequency$^\ddagger$ 	&
	Period 	&
	ID$^\dagger$	 &
	0.3-10 &
	0.3-1.0 &
	1.0-3.0  &
	3.0-10.0  & &
	0.3-10  &
	0.3-1.0  &
	1.0-3.0  &
	3.0-10.0  \\

	(mHz) 	&
	(s) 	&
	 	 &
	(keV) &
	(keV) &
	(keV) &
	(keV) & &
	(keV) &
	(keV) &
	(keV) &
	(keV)  \\
	\hline
		0.185 $\pm$0.007  & 5405 $\pm$ 193 & 2$\Omega$ & \dots & \dots &
		\dots & \dots  && 163.9 & 18.8 & 230.5	&  177.1  \\
		0.260 $\pm$ 0.006  & 3845 $\pm$ 89 & 3$\Omega$ & \dots & \dots &
		\dots & \dots  & ~ & 501.2 & 34.3 &  239.5 	&  272.8  \\
		0.320 $\pm$ 0.006  & 3125 $\pm$ 59  &  4$\Omega$ &   \dots   &
		\dots & \dots & \dots  && 278.4 & 36.7   &  139.1  &  174.9  \\
		0.366 $\pm$ 0.018   & 2732 $\pm$ 134 &  5$\Omega$ &455.1 & 156.9 &
		315.1 & 132.5  && \dots & \dots & 150.4    &  108.0  \\
		0.670 $\pm$ 0.013  & 1492 $\pm$ 29 &   9$\Omega$ & 235.0 & 80.4 &
		106.6 & 55.9  &&$\dots$ & $\dots$ & $\dots$ & \dots \\
		 0.870 $\pm$ 0.012  & 1149 $\pm$ 16 & $\omega_X-\Omega$& 411.9 &
		 180.6 & 267.5 & 56.6	  && 145.0 & 65.2 & 56.2 & 65.0 \\
	    0.949 $\pm$ 0.011  & 1054 $\pm$ 12 & $\omega_X$ & 804.8 & 391.1 &
		623.6 & $\dots$& & 351.2 & 105.9 & 444.7   & 83.2 \\ \hline
\end{tabular}

	\tablecomments{ $^\ddagger$ Frequencies are obtained from the Fourier
	spectrum of summed MOS and pn data in the energy range 0.3-10.0~keV
	(Figure~\ref{fig-ft0}). $^\dagger$ Tentative identification assuming an
	orbital period of 3.52$^{+1.43}_{-0.80}$~h (See text)}

\end{table*}

\begin{figure} \includegraphics[width=3.5in]{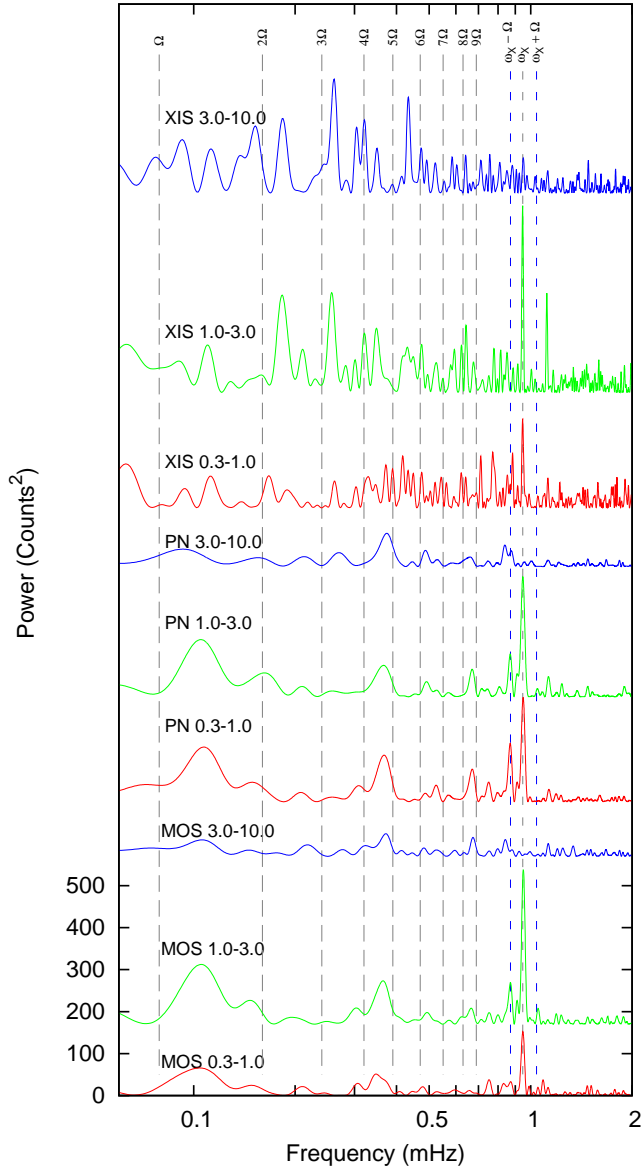}
	\caption{ Power spectra of \igr of summed \suzaku XIS,  EPIC pn and
	summed EPIC MOS light curves in different energy bands. The energy
	bands are given in the inset. A vertical offset is added to different
	power spectra for clarity.
	The different markers correspond to the harmonics of the orbital period
	of 3.52~h and the two
	orbital sidebands of the proposed spin period $\omega \pm \Omega$
	(See~Figure~2).
	}
	\label{fig-ft} \end{figure}

The energy spectra of IGR1719, discussed in a later section
(\S\ref{sec-spec}), suggest a probable soft X-ray excess in the energy
range of 0.3 to 1.0~keV. To look for a possible connection of the soft
X-ray excess with the white dwarf itself, we extracted three light curves in
the energy ranges of 0.3-1.0~keV, 1.0-3.0~keV and 3.0-10.0~keV respectively
from the source and background regions in the \xmm using the standard SAS
tools and obtained the power spectra with the help of \emph{powspec}. The
power spectra of the pn and summed MOS data are plotted
in Figure~\ref{fig-ft}. Similarly the CLEAN periodogram obtained from the
\suzaku data are also shown in Figure~\ref{fig-ft}.
From these figures, it is clearly seen that the variability
is present in all  the energy bands, though the amplitude is changing with
energy.

\begin{figure}
	\includegraphics[height=2.3in]{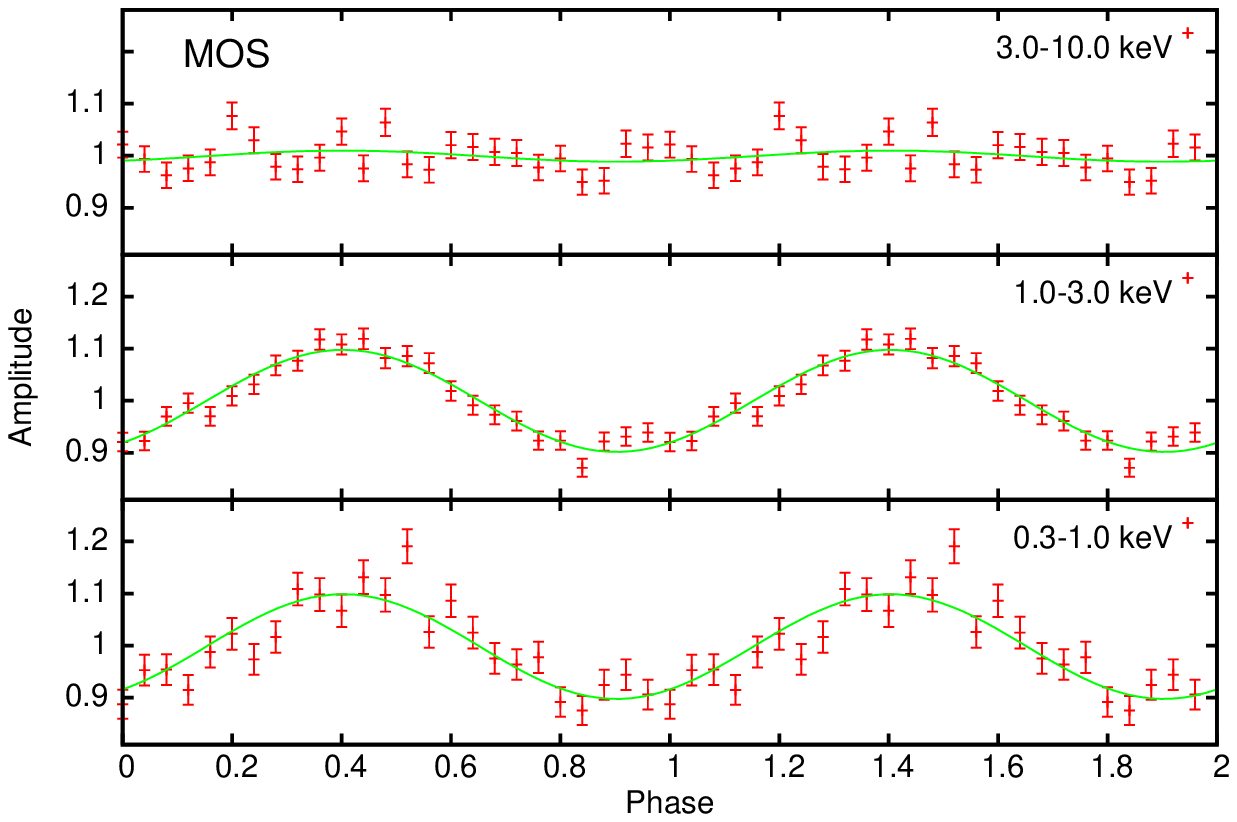}
	\includegraphics[height=2.3in]{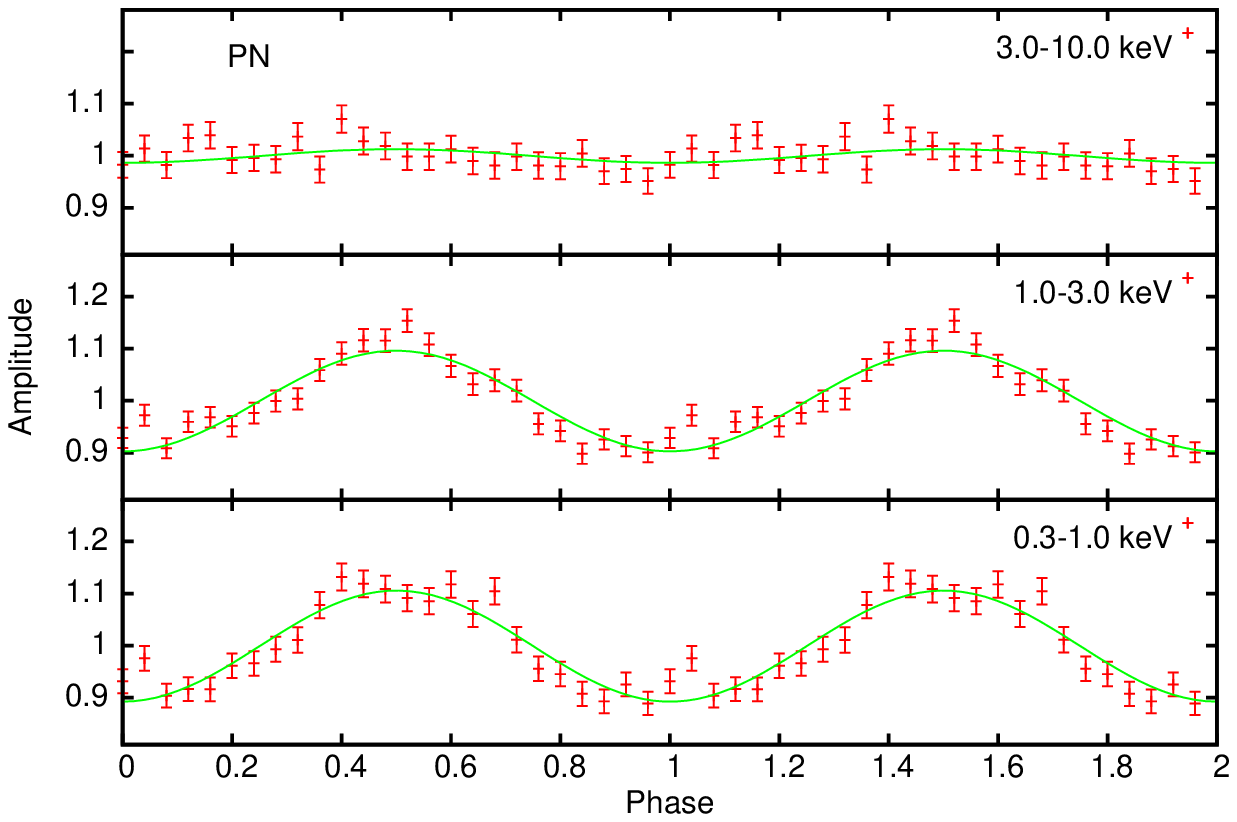}
	\includegraphics[height=2.3in]{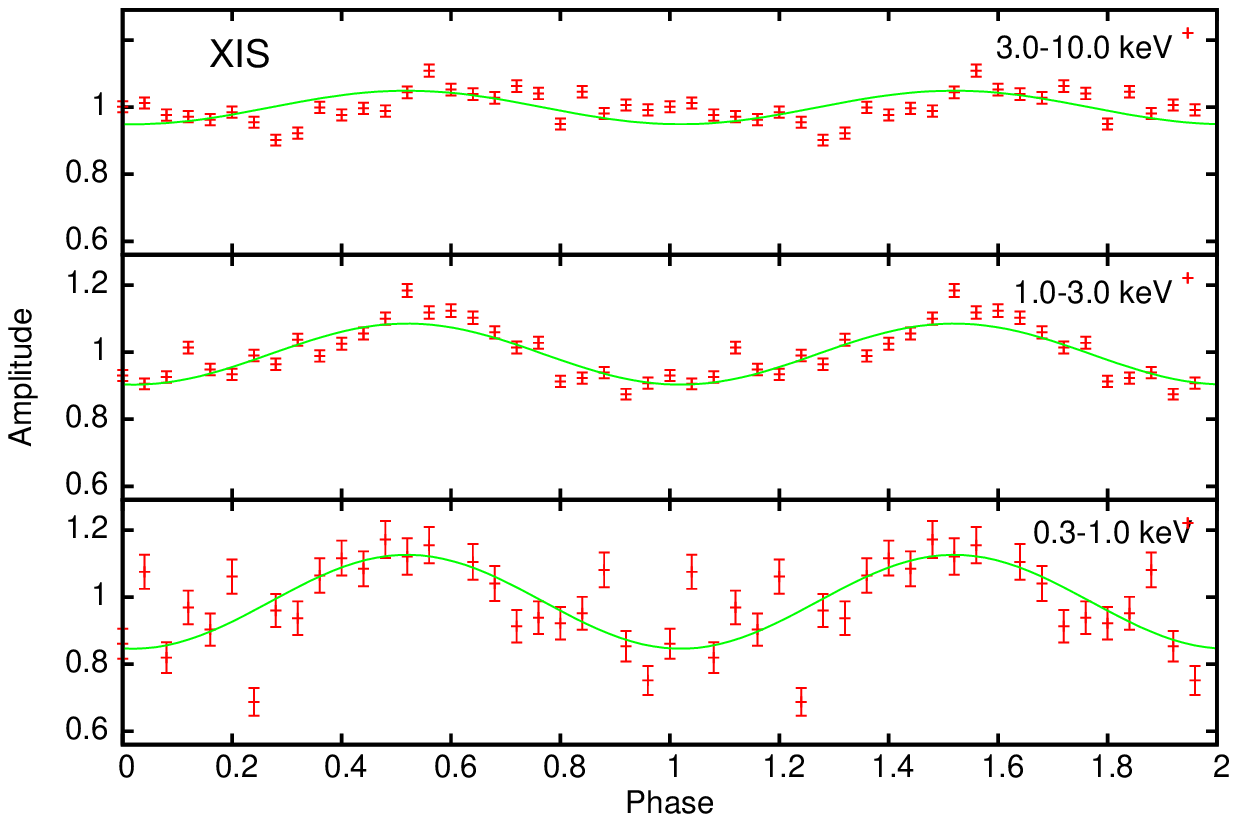}
	\caption{ Phase folded  combined MOS, pn and XIS light curves of  \igr
	folded at the period of 1053.7~s with an arbitrary phase. The instrument
	and energy bands are given in the inset. The amplitude modulation with
	the energy is evident in both the data. A simple sinusoid fitting for
	the modulation is also shown.}
	\label{fig-mosfld}
\end{figure}

Light curves of the combined EPIC MOS1 and MOS2 data were folded at the
most significant period of 1053.7~s (Fig.~\ref{fig-ft}) at different
energies, and the folded light curves are shown in Figure~\ref{fig-mosfld}
top panel.  Similarly, we did a period folding of the pn-data and XIS data 
and show it
in the middle and the lower panel of Fig.~\ref{fig-mosfld} respectively. The solid lines in all the
panels correspond to a simple sinusoid fit to the folded data. A sinusoid
variability is clearly evident in the soft energy bands viz.  0.1-1.0~keV
and 1.0-3.0~keV while it is barely seen in the hard X-ray band of
3.0-10.0~keV of \igr.

{\bf Epoch Estimation:}
To estimate the ephemeris of the most prominent peak identified in the
power spectra of IGR1719, we folded the combined XIS
lightcurve  at a period of 1053.7~s  using \emph{efold} with the integer
part of \suzaku observation (2009:09:03) as
the guess epoch. The folded lightcurve is fitted with a sine
curve to obtain the initial phase. The initial guess value of the  epoch is corrected for
the phase difference and the exercise is repeated for the combined \xmm
lightcurve using the corrected epoch obtained from \suzaku data to estimate the quadrature co-efficient. Using this method we fix the ephemeris corresponding
to the flux  maxima of \igr of the highest peak seen in Fig.~\ref{fig-ft0} as
$$ T_{max} = 55076.9993 + 0.0121975(6) E - 2.8113(4)\times 10^{-11} E^2$$

\subsection{ X-ray Spectral Analysis }
\label{sec-spec}

\begin{figure*}
	\includegraphics[width=4.2in, angle=270]{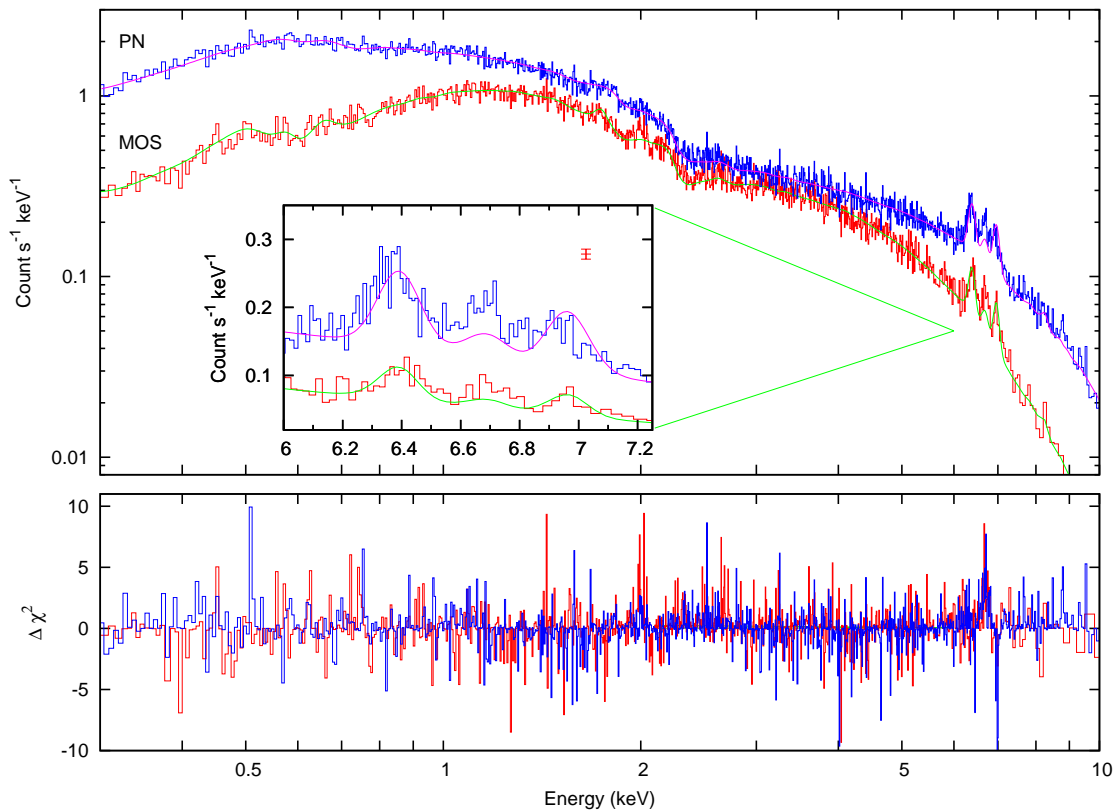}
	\caption{ Epic/pn and MOS spectra (histogram) and the best fitting
	\apec model of \igr (continuous line). The models consist of a partially
	absorbed \apec, low energy \apec and  a gaussian line. The zoomed
	spectra and the model fit around Fe-line region are shown in the inset.
	An error bar at 6.4 keV representing the typical error bar size is also
	shown in the inset.  The bottom panel shows the $\Delta \chi^2$
	contribution of data points for the best fitting model.}
	\label{fig-apec}
\end{figure*}

The background subtracted \xmm spectra  of \igr are shown as histogram in
Figure~\ref{fig-apec}.
Strong broad continuum extending to very hard X-rays along with some
emission line features can be seen in the spectra at  6.4, 6.7 and 6.9~keV.
Spectral fits were performed using \xspec  to fit  the EPIC/pn and the
combined MOS spectral data with various standard models given below.
All the models include a  common ISM absorption component (phabs). A
constant multiplicative term is also used for  the combined fits to correct
for the relative normalization difference between the MOS and the pn data.
The best fitting normalization for the pn data consistently required a factor
of 1.11 to the normalization irrespective of the models used below. The
hydrogen column density (N$_{H,ISM}$) in the \emph{phabs} model is left as
a free parameter in all the fits.

Several spectral models were tried to obtain a fit to the observed spectral
data. All the models used and their best fitting values are given in
Table~\ref{tab-apec}.  A single component bremsstrahlung or  black body
temperature models produced unacceptable fits to the \igr {}~spectra with
$\chi^2_\nu > $ 3.0. A single component  plasma emission model (\apec) also yielded an
unacceptable $\chi^2_\nu$ of 2.4 (dof: 1882) with the temperature
kT$_{apec}$ pegged at 64~keV.  The \apec plasma model however,  predicts
several emission lines due to collisional ionization in a hot thin plasma
but is not able to account for the 6.4~keV line which is most likely due to
fluorescence of cold iron.  Hence,
a Gaussian line centered at 6.4~keV and with a fixed width of 0.02~keV is
added to the \apec model resulting in an improved but still un-acceptable
$\chi^2_\nu$ of 1.99 (dof: 1879).  The \apec temperature is still pegged at
64~keV.  A partial covering absorption by neutral material seen in many
mCVs \Citep{yuasa10} was then added to the \apec + gauss model and it gives an improved
$\chi^2_\nu$ of 1.20 (dof: 1877). The best fitting value of the partial
absorber hydrogen column density is
7.4$^{+0.6}_{-0.6}\times$10$^{22}$~cm$^{-2}$ with a covering fraction of
44$\pm$2 per cent and the \apec temperature is 19.1$^{+1.0}_{-1.6}$~keV.

Though the \apec + gauss models  give a better fit and a
realistic shock temperature, there
is still an unexplained excess seen around 0.5~keV, in the form of
several unresolved lines.
Therefore, a low temperature \apec component was found to be more suitable
than a blackbody
to improve the chi-square. Spectral fits with an additional blackbody
component,  usually present in some IPs, were also tried but yielded no
improvement in the chi-square.
To account for the excess emission around 0.5~keV, we tried two additional models:
(a) a low energy \emph{apec}, and (b)  a narrow Gaussian,
which gave $\chi^2_\nu$ values of 1.02 (dof: 1875) and 1.04 (dof: 1874)
respectively (see, Table~\ref{tab-apec}). The best fitting N$_{H,ISM}$ value
converges to $\sim$1.12$\pm$0.02$\times$10$^{21}$~cm$^{-2}$ for all the
models.

\begin{table*}
	\caption{Spectral parameters of \igr for different
	models. The temperature kT corresponds to the temperature of
	\apec or the low temperature of \emph{MKCFLOW} components.
	The errors indicate the 90 per cent confidence values.}
\label{tab-apec}
\begin{tabular}{cccccccc}
\hline
Model  & pcf(apec) & pcf(apec)  & pcf(apec) &
pcf(apec)& pcf(mkcflow)  &
pcf(mkcflow) & pcf(mkcflow) \\
Parameter &  & +Fe line & +Fe line+ga & apec + Fe line & + Fe line &
 +Fe line+ga &  +apec+Fe line \\ \hline
 $\chi^2$ (dof) & 1.35 (1879) & 1.20 (1877) & 1.04 (1874) & 1.02 (`875) &
 1.16(1876) & 1.00 (1874) & 0.97 (1875) \\[2mm]

 $^a$N$_{H,ISM}$ & 80 $\pm$ 2 & 80 $\pm$ 2 & 98 $\pm$ 4 & 112 $\pm$ 2 &
			   76 $\pm$2 & 88$\pm$3 & 112 $\pm$ 2	\\
 $\times10^{19}$ cm$^{-2}$ \\[2mm]

 $^b$N$_{H,pcf}$  &  8.4 $^{+0.7}_{-0.6}$  & 7.4$^{+0.6}_{-0.5}$ &
 8.4$^{+0.6}_{-0.6}$ &8.7$^{+0.7}_{-0.7}$ &6.2$^{+0.5}_{-0.5}$
 &6.8$^{+0.6}_{-0.5}$ & 7.3$^{+0.7}_{-0.6}$ \\
 $\times10^{22}$ cm$^{-2}$ \\[2mm]

 $^c$pcf$_{frac}$  &  46 $\pm$ 2 & 44 $\pm$ 2 & 46$^{+1}_{-4}$ & 46$^{+2}_{-1}$ &
 40 $\pm$ 2 & 41 $^{+1}_{-2}$ & 40$^{+1}_{-2}$
 \\
 \% \\[2mm]

 Abundance & 0.51 $^{+0.08}_{-0.07}$ & 0.56 $^{+0.08}_{-0.07}$ & 0.52 $^{+0.08}_{-0.07}$ & 0.52 $^{+0.08}_{-0.07}$ &
0.92 $^{+0.08}_{-0.07}$ & 0.85 $^{+0.08}_{-0.07}$ & 0.84 $^{+0.08}_{-0.07}$ \\
 ($\odot$) \\[2mm]

 $^d$kT$_{apec/mkcf}$ & 
 19.8 $^{+1.0}_{-1.7}$ & 19.1 $^{+1.0}_{-1.6}$ & 18.7 $^{+1.9}_{-3.3}$ & 16.0 $^{+1.0}_{-1.1}$ &
 4.6 $^{+0.8}_{-0.9}$ & 4.4 $^{+0.8}_{-0.8}$ & 3.9 $^{+0.7}_{-0.8}$ \\
 (keV) \\[2mm]

 kT$_{apec2}$ & 
 \dots & \dots & \dots & 176.4$^{+6.2}_{-7.7}$ &  
 \dots & \dots & 174.8$^{+4.8}_{-7.6}$    \\
 (eV) \\[2mm]

 Eq Width & \dots & 134.6 & 132.4 & 126.5 & 128.6 & 126.5 & 138.7 \\
 Fe~K$\alpha$(eV) \\[2mm]

 Eq Width & \dots & \dots & 47 & \dots & \dots & 49 & \dots \\
 0.58~keV (eV) \\[2mm]

 $^e$X-ray Flux & 3.16 &  3.12 & 3.11 & 3.09 & 3.19 & 3.15 & 3.12 \\
 0.3-10 keV \\
 ($\times$10$^{-11}$ erg cm$^{-2}$ s$^{-1}$) \\[2mm]

 $^f$L$_{X}$ &  4.57 & 4.52 & 4.50 & 4.47 & 4.62 & 4.56 & 4.52
 \\
 0.3-10 keV \\
 $\times10^{31}$ (erg s$^{-1}$) \\[2mm]
 \hline
\end{tabular}

\tablecomments{
$^a$N$_{H,ISM}$ = ISM column density,
$^b$N$_{H,pcf}$ = Partial covering absorber column density,
$^c$ pcf = partial covering fraction,
$^d$ Only $kT_{low}$ values are given for \emph{mkcflow} component,
$kT_{high}$ is un-constrained at  $>$74 keV. 
$^e$X-ray flux as measured with the MOS detectors.
$^f$X-ray luminosity assuming a distance of 110~pc.}

\end{table*}

\begin{figure*}
	\includegraphics[width=4.2in,angle=270]{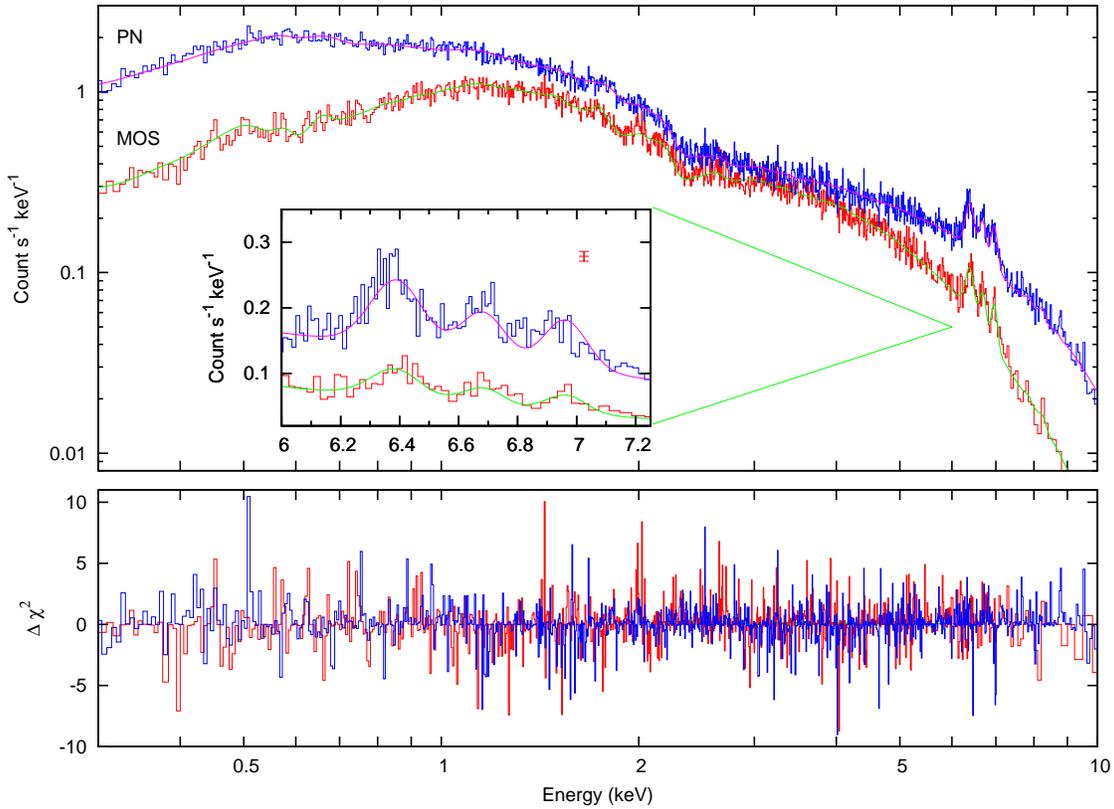}
	\caption{ Same as Figure~\ref{fig-apec} for MKCFLOW models}
	\label{fig-mkcflow}
\end{figure*}

The best fitting was thus obtained for a model consisting of a partially
absorbed \apec component + a low energy \apec component + a Gaussian at
6.4~keV. The best fitting values for the partial absorber were: column
density N$_H$ $\approx$ 1.12$\pm$0.02$\times$10$^{-21}$~cm$^{-2}$ with a
covering fraction of 46$^{+1}_{-4}$~per~cent.  The best fitting
temperatures of the two \apec components are 16.0$^{+1.0}_{-1.1}$~keV and
176.4$^{+6.2}_{-7.7}$~eV respectively for a common elemental abundance of
0.52$^{+0.07}_{-0.06}$.  The error bars quoted here and everywhere in this
paper are for a 90~per~cent confidence limit for a single variable
parameter. The best fitting \apec model to the observed MOS and pn spectra
and their contribution to chi-square of the fit  are shown in
Figure~\ref{fig-apec}.  Though this fit by two component \emph{apec} models
gave $\chi^2_\nu$ close to 1.0, it was unable to fit the plasma emission
line at 6.7~keV presumably due to the very high or very low temperatures of
the plasma that were fitted.  This suggests that a multi-temperature plasma
model may be more appropriate than a single temperature \emph{apec} model
to reproduce the \igr spectra.  \citet{mukai03} and \citet{pandel05} showed
that the cooling flow model MKCFLOW available in \xspec is very successful
in reproducing the \emph{Chandra} and \xmm spectra of several mCVs.  We,
therefore, fitted a model consisting of partially absorbed MKCFLOW and a
Gaussian line at 6.4~keV and obtained a $\chi^2_\nu$ of 1.16 for 1876
degrees of freedom. The addition of a second Gaussian at 0.52~keV or a low
kT (175$^{+5}_{-8}$~eV) \apec component further improved the fit to give a
$\chi^2_\nu$ of 1.00 (dof: 1874) and 0.97 (dof: 1875) respectively. The
best fitting value of the low temperature parameter in the MKCFLOW model is
3.9$^{+0.7}_{-0.8}$~keV while the high temperature value is not constrained
but pegged at the maximum model temperature of 80~keV.  The best fitting
N$_{H,ISM}$ value at 1.12$\pm2\times$10$^{21}$~cm$^{-2}$ is the same as in
the earlier two temperature  \apec model fits. The best fitting values for
the partial absorber column density at
7.3$^{+0.7}_{-0.6}\times$10$^{22}$~cm$^{-2}$ and covering fraction of
40$^{+1}_{-2}$~per~cent  are consistent with the values obtained using two
temperature \apec models previously (see the first four models in
Table~\ref{tab-apec}).  Similarly, the temperature of the additional \apec
component kT$_{apec}$ at 174.8$^{+4.8}_{-7.6}$~eV matches with the
temperature derived with two temperature \apec models.  The best fitting
MKCFLOW model fit is shown in Figure~\ref{fig-mkcflow} and the spectral
parameters of the model are listed in Table~\ref{tab-apec} (last
3 models). It is also to be noted that both the 6.7 and 7.0~keV plasma
emission lines are now fitted (Fig.~\ref{fig-mkcflow}).

Assuming a distance of 110~pc for \igr \Citep{masetti06}, we estimate the
X-ray luminosity from the flux in 0.3-10.0~keV energy band for all the
models. The X-ray flux and estimated X-ray luminosity in the energy range
of 0.3-10.0~keV are listed in Table~\ref{tab-apec}.

We calculated the un-absorbed soft (F$_s$) and hard (F$_h$) X-ray flux in
the energy band of 0.2-12.0~keV from the low kT \apec model and MKCFLOW
model respectively. The fluxes are
F$_s$~=~1.26$\pm$0.12$\times$10$^{-13}$~erg~cm$^{-2}$ and
F$_h$~=~4.92$\pm$0.11$\times$10$^{-12}$~erg~cm$^{-2}$ respectively. The
softness ratio is then calculated using the relation F$_s/4F_h$
\Citep{evans07} as 0.0064 (-2.19$\pm$0.03 in $\log_{10}$ scale).

\subsection{Phase Resolved Spectroscopy for the dominant period }
\label{sec-phi}

The spectral fits discussed in \S\ref{sec-spec} show that a partial
covering absorber can explain the \xmm spectra of IGR1719, suggesting a
varying absorption along  the line of sight. For a better understanding of
the spectral behaviour of \igr as a function of the phase for the dominant
period present in Fig.~\ref{fig-ft0} \& \ref{fig-ft}, we analyzed the
spectra extracted for five equal phase intervals of 0.2. We extracted the
\xmm spectra at different  phases of the most dominating period of 
1053.7~seconds (see \S\ref{sec-timing}).

With the \emph{phacalc} tool available in SAS analysis suite, the phases
corresponding to the frequency $\omega_X$ (0.949~mHz) of all the events
were calculated for the MOS1, MOS2 and pn detectors  using
MJD~55077.00663866 as the epoch.  The tool adds a `PHASE' column  to
the events file. The spectra corresponding to different phases with a
uniform phase bin of 0.2 with $\Phi_0$ = 0.0, 0.2, \dots were extracted
using phase value as an additional criterion to the standard selection
criteria used earlier (see, \S\ref{sec-spec}).  The MOS1 and MOS2 spectra
were combined together to improve the statistics.  Since we are interested
only in the partial absorber component, parameters of other models like the
N$_{H,ISM}$, temperature of the \apec component  were fixed at the values
shown in Table~\ref{tab-apec} while varying only their normalization
values.

\begin{figure}
	\includegraphics[angle=-90,width=3.3in]{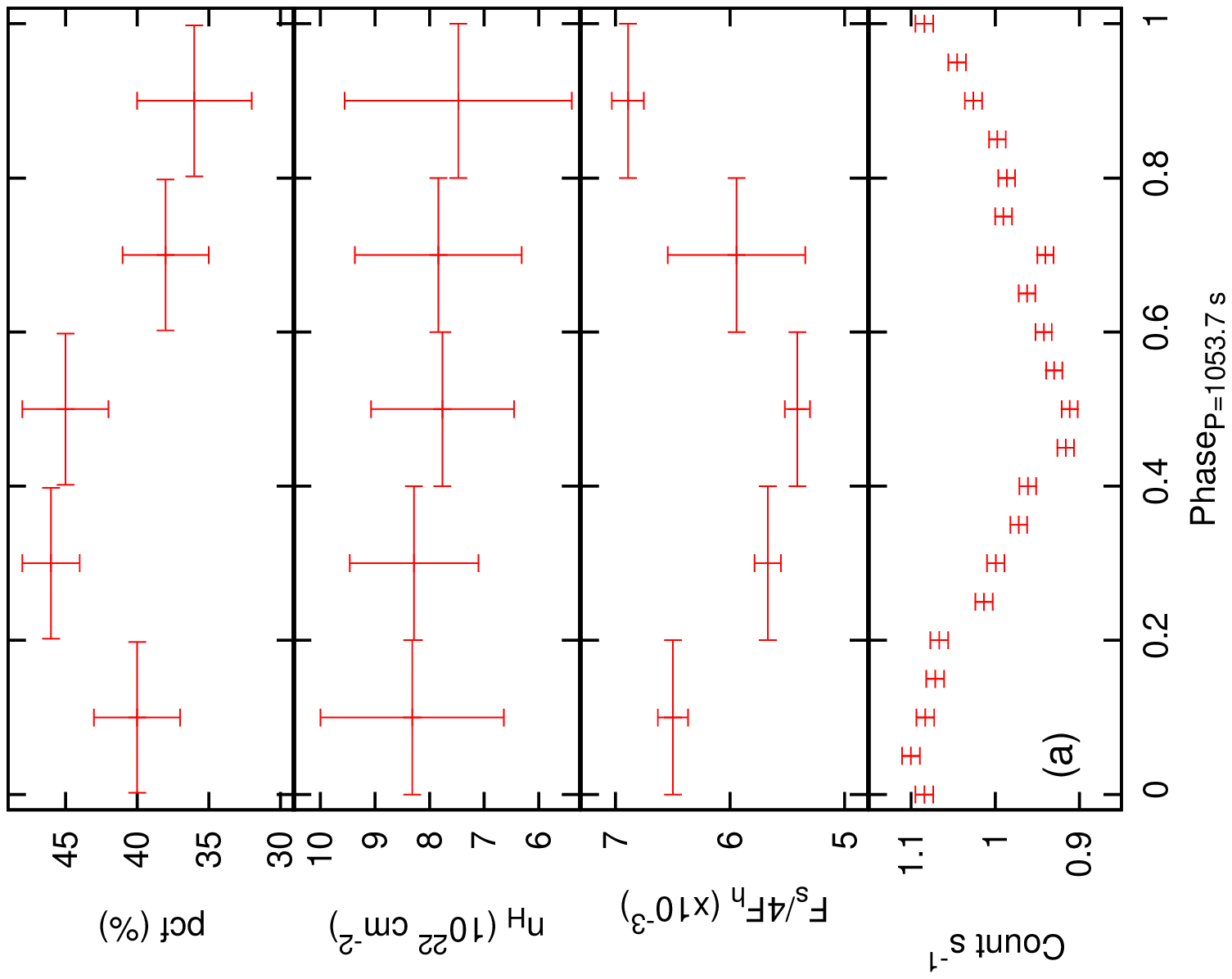}
	\includegraphics[angle=-90,width=3.3in]{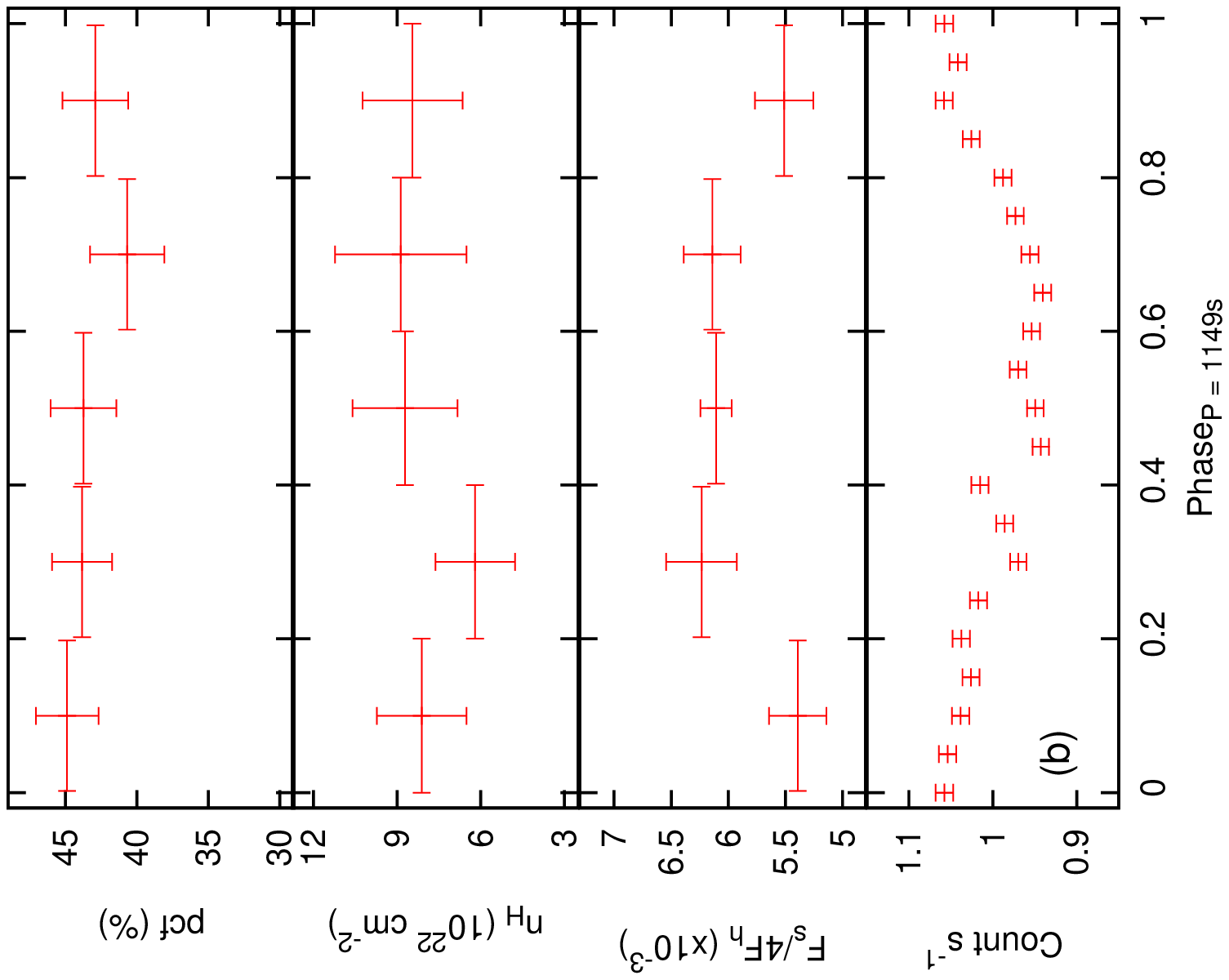}
	\caption{The partial covering absorber per cent and the  N$_{H,pcf}$
	parameter values as a function of phase of the dominant period (P$_X$ =
	1053.7 s ) seen in \igr X-ray data along with the softness ratio of
	\igr defined as F$_s/4F_h$ (a).  Also shown is \xmm lightcurve in the
	energy band 1.0-3.0~keV folded at the spin period and scaled to match
	the absorber fraction is also shown for comparison (stars). The right
	Figure  corresponds to the phase of P=1149.4 s (b).}
	\label{fig-phi}
\end{figure}

The spectral fittings were performed on the combined MOS spectra with \\
apec+gauss+pcfabs(MKCFLOW) model.  The  partial covering absorber fraction,
the equivalent hydrogen density of the column  as a function of spin phases
of \igr are shown in Figure~\ref{fig-phi}a along with the folded \xmm
0.3-1.0~keV light curve profile for a comparison.  We calculated the
un-absorbed flux in the soft (F$_s$) and hard (F$_h$) components in the
energy range 0.2-12.0~keV separately and obtained the softness ratio
F$_s$/4F$_h$.  The softness ratio as a function of phase is shown in
Fig.~\ref{fig-phi}a.  An anti-correlation between the X-ray intensity and
the partial absorber fraction, and a correlation between  the softness
ratio and the X-ray intensity can be seen in Fig.~\ref{fig-phi}a.  The N$_H$
value of the partial absorber, however, appears to be constant within the
error bars.

A similar study was performed with respect to the phases of the second
dominant X-ray period at 1149.4~s also and plotted in
Figure~\ref{fig-phi}b. From the figure, it is evident that the softness
ratio seems to show anti-correlated variation with the phase whereas the
partial covering fraction remains constant.  The behaviour of the softness
ratio is not consistent with the X-ray intensity variations seen in Figure
4.

\section{ Discussion }

The temporal analysis of \xmm and \suzaku data of IGR1719 shows the
presence of several power peaks at  frequencies below 1~mHz.  The
continuous \xmm data  helps in identifying the  periodic variability 
of the system un-ambiguously.  The dominant frequency here corresponds to the
period 1053.7$\pm$12.2~s (P$_X$) in the soft  energy bands viz.
0.3-1.0~keV and 1.0-3.0~keV.  There is only a hint of this period in the
hard X-rays, with a much reduced amplitude.  Another peak at a period of
1149.4$\pm$15.8~s near the main frequency is also seen in both the MOS and
pn power spectra.  This period is close to the previously reported spin
period of 1139.55$\pm$0.04~s by \Citet{pretorius09} based on high speed
optical photometry  and attributed to the spin of the white dwarf.
\Citet{pretorius09} also reported an orbital period of 4.005$\pm$0.006~h
for \igr in the time-resolved spectroscopy data  and noted the presence of
some  power around the beat frequency between orbital and then assumed spin
period. However, they also pointed out  that, with the available optical
data, they were unable to distinguish between the spin period and the beat
period.  It is likely, therefore, that the previously reported spin period
might actually be the sideband period due to beating between the orbital
and spin periods (orbital side-band).  If we assume that the dominant peak
seen here at 1053.7$\pm$12.2~s in the X-ray light curves actually
corresponds to the spin frequency ($\omega_X$) of the white dwarf and
1149.4$\pm$15.8~s period is the orbital side band ($\omega_X - \Omega$),
then this would  imply an orbital period ($P_{orb}$) of
3.52$^{+1.43}_{-0.80}$~h. This orbital period and its harmonics are
marked as long dashed lines in Fig.~\ref{fig-ft}, and the spin frequency
and orbital beat frequencies of the spin frequency are marked as short
dashed lines. Our estimation of P$_{orb}$ = 3.52$^{+1.43}_{-0.80}$~h based
on the above assumption  being in agreement with the orbital period of
4.005$\pm$0.006~h reported by \citet{pretorius09}. 
Table~\ref{tab-pers} lists the observed optical and X-ray periods and their
plausible identifications along with the estimated orbital periods. We also
calculate the expected beat periods assuming the dominant X-ray period as
the spin and the orbital period as obtained by \Citet{pretorius09} and list
them in  Table~\ref{tab-pers} for comparison.

\begin{table}
	\caption{ Estimation of orbital period of \igr by assuming the
	dominant X-ray period  as the spin period (first row) and the optical
	period as the spin period (second row).  
	Calculated beat periods corresponding
	to 	the dominant X-ray peak and the reported orbital period of 4.005~h
	\citep{pretorius09}
	are also listed in the third row. 
	}
	\label{tab-pers}
	\begin{tabular}{cccc}
		\hline
		Spin Period  &  \multicolumn{2}{c}{Beat Periods}  &  Orbital Period \\ \cline{2-3}
		P$_{spin}$	  & P$_{\omega - \Omega} $ & P$_{\omega + \Omega}$ & P$_{orb}$ \\	
	(s)   &  (s)  & (s) & (h) \\ 
	\hline
	1053.7 $\pm$ 12.2 (X)	 & 1149.4 $\pm$ 15.8 (X) &
	\dots  &
	3.52$^{+1.43}_{-0.80}$ \\ 
	1139.55 $\pm$ 0.04 (o)	& \dots & 1053.7 $\pm$
	12.2 (X) &3.89$^{+0.69}_{-0.53}$  \\ 
	  
	1053.7 $\pm$ 12.2 (X)	&  1136.8$\pm$ 14.3$^a$ 	 &
	981.9 $\pm$ 10.5$^a$	 & 4.005 $\pm$ 0.006$^b$  \\ 
	\hline
		\end{tabular}

		\tablecomments{
		X,o - observed periods in X-ray and optical data,
		Optical period value is taken from \Citet{pretorius09},
		a - Calculated using reported orbital period of \igr, 
		b - Obtained from radial velocity measurements \Citep{pretorius09} 
		}
\end{table}


According to \citet{wynn92}, the power spectrum of an IP with a simple
disc-less geometry is dominated by signals at spin frequency ($\omega$),
beat frequency ($\omega - \Omega$) and second beat frequency (2$\omega -
\Omega$) components with the condition that one of the latter should always
be present, whereas the disc accreting systems show only the spin frequency
component ($\omega$).  Therefore, the presence of a dominating power at
spin period ($P_x$=1053.7~s) and a significant power at the orbital
sideband at 1149.4~s in the \igr X-ray power spectra suggests a predominant
contribution from the disc-less accretion component in \igr 
making \igr a member of the rare disc-less accreting IPs.  X-ray pulsations
at the orbital beat period are seen when the accreting material is directly
coupled with the magnetic field lines. The power spectra of \igr are
similar to RX J1803 power spectra of \xmm lightcurves though the modulation
at the beat period is significantly less compared to the spin period
modulation \Citep{anzolin08}.
\Citet{wynn92} also suggest that in a disc-less accretion system, the power
in the spin period decreases with the energy, whereas the power in the beat
period increases with energy. This is attributed to the absorption of low
energy photons by the accretion column itself. In the intermediate polars,
the highest energy X-rays are produced by shocks above the white dwarf
surface and hence are less prone to absorption by the accretion column
resulting in the decrease of modulation.  The softer X-rays are emitted
closer to the white dwarf surface than the harder X-rays are, and are thus
more prone to absorption by the accretion column and hence, resulting in
the increase of amplitude modulation.
The \igr power spectra obtained in different energy bands show
that the amplitudes of modulation is decreasing with energy
(Fig.~\ref{fig-ft}).
The near absence of any
significant power near the spin period or the orbital side band in the hard
X-ray power spectrum also suggests that the hard X-rays originate at a much
higher altitude in the accretor.
Combining this zero power  at the spin period in the hard X-rays ($> 3$keV),
but a significant power in the beat period at 1149.4~s along with power in
orbital period and its harmonics in the hard X-rays further supports the
disc-less accretion \Citep{wynn92}.

A partial absorber in front of a multi-temperature plasma in a  cooling
flow model best explains the \xmm spectra of IGR1719. The multi-temperature
plasma produced in the shock region just above the white dwarf is thus seen
through two absorbers, an ISM component (N$_H$ =
1.12$\times$10$^{21}$~cm$^{-2}$) and a thick partial covering absorber
(N$_H$ = 7.3$\times$10$^{22}$~cm$^{-2}$).
The estimated value of the partial covering absorber fraction matches with
the fraction obtained by \citet{yuasa10} for \igr using \suzaku data.  The
spectra of \igr indicate the presence of a very high temperature for which
we could only obtain a lower limit.  \Citet{yuasa10} reported the mass of
white dwarf to be 1.03$^{+0.24}_{-0.22}$~M$_\sun$.
We estimated the maximum shock temperature corresponding to this mass as
64.6$^{+18.5}_{-7.6}$~keV using the relation, $$ T_{shock} = \frac{3}{8}
\frac{GM_{wd} \mu m_H}{kR_{wd}} $$ where the radius R$_{wd}$ is calculated
from the  white dwarf mass-radius relation given by \citet{nauenberg72} is
5.24$^{+1.68}_{-2.05}\times$10$^8$~cm. The calculated shock temperature of
the \igr explains the un-bound high temperature values obtained from  the
cooling flow spectral models.  The Fe abundance in the multi temperature
plasma is very close to solar abundance and reproduces the observed Fe
lines at 6.7 and 7.0~keV completely.

The X-ray flux obtained from different spectral models are very similar in
the value.  For the best fitting apec+MKCFLOW model, the flux  in
0.3-10.0~keV energy band is estimated to be 3.1$\times$10$^{-11}$
erg~cm$^{-2}$~s$^{-1}$ which yields an X-ray luminosity of
4.5$\times$10$^{31}$ erg~s$^{-1}$.  Using the relation L=GM\.M/2R and
taking the mass of the white dwarf as 1.03~M$_\sun$ \Citep{yuasa10} and a
radius of 5.24$\times$10$^8$~cm calculated from the \citet{nauenberg72}
white dwarf mass-radius relation, we obtain a mass accretion rate of
5.5$\times$10$^{-12}$~M$_\sun$~yr$^{-1}$.

A growing number of IPs show a soft X-ray excess with blackbody
temperatures less than 100~eV (\Citealt{evans07}, \Citealt{anzolin08},
\Citealt{anzolin09}, \Citealt{martino08}).  Though the \xmm spectra of \igr
show a soft excess, the residuals showed several un-resolved lines around
0.5~keV energy, and the spectral fits with an additional blackbody
component did not improve in the chi-square.  Alternatively a low temp
\apec component with a kT of 175~eV did improve the chi-square, and was
found to be more suitable.  This is also consistent with the improvement
seen using continuous temperature models like MKCFLOW.  The kT$_{apec}
\sim$175~eV is significantly higher than the typical blackbody temperatures
(30-120~eV) of other soft IPs.

\Citet{evans07} suggest that the main reason for some IPs to show soft
excess is that the accretion curtains do not hide the accretion footprints.
The soft X-ray excess in \igr thus suggests a highly inclined magnetic axis
in the system.  We calculated a softness ratio of 0.0064 from the
unabsorbed flux of \igr which matches the softness ratio of V2400~Oph  within
one sigma error \Citep{evans07}. This is interesting, that although a soft
X-ray component with a ratio similar to that in  other IPs exists in
IGR1719, but a simple black body model typically used for the soft excess
is unable to explain the soft excess seen in IGR1719.

The multi-temperature plasma models do not reproduce the observed 6.4~keV
Fe fluorescent line, thus  requiring the addition of a Gaussian component.
The observed fluorescent line is compatible with the emission from
comparatively cold iron in the ionization states Fe I to Fe XVII
\Citep{andre96} and does not show any significant broadening.  This line
feature in the X-ray spectra is seen in several IPs and polars
(\Citealt{hellier04}, \Citealt{landi08}) and usually understood as due to
reflection from cold matter.

From the spin phase resolved spectroscopy ($\S$\ref{sec-phi}), we found
that the variation in the partial covering fraction (pcf) of the absorber
is anti-correlated with the X-ray intensity whereas  the softness ratio
(F$_s/4F_h$) varies in phase with the X-ray intensity. In an IP with an
accretion curtain, self absorption can cause such a variation in the X-ray
intensity. When the curtain is maximally in the line of sight, the X-ray
intensity will be minimum due to the maximum absorption, and the light
intensity will be the maximum when either the curtain moves away from the
line of sight or when it thins down.  Similarly, more of the soft X-ray
photons assumed to be from the accretion footprint will be absorbed if the
accretion curtain is in the line of sight resulting in  the reduction of
softness ratio, and the softness ratio will increase when the curtains do
not hinder the accretion footprint.  Hence, the soft excess appears to be
related to the pcf only, indicating that the phase modulation in the soft
x-rays is due to  varying coverage by the accretion curtains towards the
line of sight.

In our discussion, we have assumed that the X-ray period of 1053.7~s 
in the XMM-Newton data is the spin period of the white dwarf and
the period at 1149.4~s is the orbital side-band. But, we cannot rule
out the plausibility of 1149.4~s being the spin period and 1053.7~s being
the side-band.  Also, the 1149.4~s period almost matches with  1139.6~s period
seen in the optical photometry data \citep{pretorius09} of \igr and is
reported to be due to spin of the white dwarf.  If true, spin at 1149.4~s
and orbital side-band at 1053.7~s  would
imply a retrograde white dwarf rotation in \igr similar to IGR~J1817 and
XSS~J0056 \citep{bernardini12}.

Generally the spin modulation of an intermediate polar is attributed to a
combination of varying photo electric absorption and self absorption
between the observer and the X-ray emitting regions. Hence, we are expected
to see an anti-correlation of partial covering fraction with the spin. A
near constant partial covering fraction with the phase of 1149.4~s period
and an anti-correlation with the phase of 1053.7~s period thus support
our earlier assumption of 1053.7~s as the white dwarf spin and 1149.4~s
as the orbital side-band.

\section{ Conclusions : }

Our timing analysis of the archival \xmm and \suzaku data of IGR1719 shows
a dominant frequency at 0.949$\pm$0.011~mHz corresponding to a period of
1053.7$\pm$12.2~s and another close period at 1149.4~s. We propose that the
period at 1053.7~s is the spin period of the white dwarf, and the period at
1149.4~s is the orbital sideband, and thus we estimate the orbital period
as 3.52$^{+1.43}_{-0.80}$~h. We also provide an  ephemeris of this spin
period.  Though the results of the phase resolved spectroscopy performed at
the two periods 1053.7~s and 1149.4~s favours the proposed period at
1053.7~s as the white dwarf spin period, polarimetric detection of the spin
period of \igr is needed to confirm the spin period. The power spectra
obtained in different energy bands suggest that \igr is a candidate for the
rare group of disc-less IPs.  

A partially and fully absorbed cooling flow model satisfactorily explains
the observed spectra of IGR1719.  The partial covering fraction varies
between 0.34 at the spin minima to 0.46 at the spin maxima indicating a
geometrical change in the absorber with spin phase. We also report a soft
X-ray excess which can be modelled with a low kT \apec model.  We detect a
6.4~keV fluorescent line which can be attributed to the emission from cold
iron in the ionization states up to Fe~XVII from the white dwarf surface.
From the observed flux in the 0.3-10.0~keV energy range, we estimate a mass
accretion rate of 5.5$\times$10$^{-12} M_\sun$~yr$^{-1}$.

From the temporal and spectral analysis the image of \igr that emerges is
of a disc-less accreting system where the matter from the secondary  is
threaded along a magnetic pole as a stream resulting in modulations of
X-rays at the orbital beat period. The shocks formed above the white dwarf
surface give a multi temperature spectra with a shock temperature above
64~keV. The soft X-ray excess seen in \igr may be emitted by the unhindered
accretion footprint. The site for origin of the fluorescent line emission
is not clear in the system.

~\\~

\noindent\emph{Acknowledgements:} This research has made use of data
obtained from the High Energy Astrophysics Science Archive Research Center
(HEASARC), provided by NASA's Goddard Space Flight Center.  We would like
to thank the anonymous referee for the insightful remarks that improved the
manuscript.


\bibliographystyle{apj1b}

\end{document}